\documentclass[trackchanges]{aastex701}
\usepackage[utf8]{inputenc}
\def\mjybm{{\rm mJy\,beam^{-1}}}

\begin{document}

\title{A tool of Hierarchical cOre ideNtification and Kinematic property AssIgnment ({\sc honkai}) for Dense Cores}

\author[orcid=0009-0007-3201-9100,sname='Liu']{Jiawei Liu}
\affiliation{State Key Laboratory of Radio Astronomy and Technology, National Astronomical Observatories, Chinese Academy of Sciences, Beijing 100101, China}
\affiliation{University of Chinese Academy of Sciences, Beijing 100049, People’s Republic of China}
\email[show]{liujw@bao.ac.cn}  

\author[0000-0003-4659-1742]{Zhiyuan Ren}
\affiliation{State Key Laboratory of Radio Astronomy and Technology, National Astronomical Observatories, Chinese Academy of Sciences, Beijing 100101, China}
\affiliation{CAS Key Laboratory of FAST, NAOC, Chinese Academy of Sciences, Beijing, China; University of Chinese Academy of Sciences, Beijing, China}
\email[]{renzy@nao.cas.cn}

\author[0000-0003-3010-7661]{Di Li}
\affiliation{New Cornerstone Science Laboratory, Department of Astronomy, Tsinghua University, Beijing 100084, China}
\affiliation{State Key Laboratory of Radio Astronomy and Technology, National Astronomical Observatories, Chinese Academy of Sciences, Beijing 100101, China}
\email[]{dili@tsinghua.edu.cn}

\author[orcid=0000-0002-2738-146X,sname='Xie']{Jinjin Xie}
\affiliation{State Key Laboratory of Radio Astronomy and Technology, Xinjiang Astronomical Observatory, Chinese Academy of Sciences, 150 Science 1-Street, Urumqi, Xinjiang 830011, China}
\email[]{jinjinxie@nao.cas.cn}

\author[orcid=0000-0001-8509-1818,sname='Fuller']{Gary A. Fuller}
\affiliation{Jodrell Bank Centre for Astrophysics, Department of Physics \& Astronomy, The University of Manchester, Manchester M13 9PL, UK}
\affiliation{I. Physikalisches Institut, University of Cologne, Z\"ulpicher Str. 77, 50937 K\"oln, Germany}
\email[]{g.fuller@manchester.ac.uk}

\author[0000-0002-3411-9654]{Yuchen Xing}
\affiliation{State Key Laboratory of Radio Astronomy and Technology, National Astronomical Observatories, Chinese Academy of Sciences, Beijing 100101, China}
\affiliation{University of Chinese Academy of Sciences, Beijing 100049, People’s Republic of China}
\email[]{xingyc@bao.ac.cn}

\author[0000-0001-8550-0095]{Xin Lyu}
\affiliation{State Key Laboratory of Radio Astronomy and Technology, National Astronomical Observatories, Chinese Academy of Sciences, Beijing 100101, China}
\affiliation{University of Chinese Academy of Sciences, Beijing 100049, People’s Republic of China}
\email[]{lvxin@bao.ac.cn}

\author[0000-0001-5950-1932]{Fengwei Xu}
\affiliation{Max Planck Institute for Astronomy, Heidelberg, Germany}
\email[]{fengwei@mpia.de}

\author[0000-0001-8923-7757]{Chen Wang}
\affiliation{Institute of Astronomy and Information, Dali University, Dali, 671003, China}
\email[]{wangchen@pmo.ac.cn}

\author[0000-0002-5927-2049]{Fanyi Meng}
\affiliation{New Cornerstone Science Laboratory, Department of Astronomy, Tsinghua University, Beijing 100084, China}
\email[]{mengfanyi@mail.tsinghua.edu.cn}

\author[0000-0002-9151-1388]{Sihan Jiao}
\affiliation{State Key Laboratory of Radio Astronomy and Technology, National Astronomical Observatories, Chinese Academy of Sciences, Beijing 100101, China}
\email[]{sihanjiao@nao.cas.cn}

\begin{abstract}

Infrared dark clouds (IRDCs) contains cold dense gas at the earliest stage of massive star and cluster formation. In studying the IRDCs, a universal and fundamental task is to resolve their internal hierarchical structures. Various packages and algorithms were developed for this purpose, but with most of them mainly focused on certain individual steps in data processing. In this work, we build a more automatic procedure for multi-band structure measurement {\sc honkai} (Hierarchical cOre ideNtification and Kinematic property AssIgnment), which can resolve the elemental components including cores and clumps, disentangle the velocity components in spectral data, measure their physical properties, and generate a catalogue for all the measured properties. We use {\sc honkai} for a joint study towards three IRDCs observed in 850 $\micron$ dust continuum with James Clerk Maxwell Telescope (JCMT) and the $^{13}CO$ $(1-0)$ data cube with the Purple Mount Observatory 14-m telescope. 193 dense cores in 16 clumps are identified. As major dynamical properties, a large amount of the cores (136 out of 193) are measured to have large virial ratio of $R_{\rm vir}>1$, but their mass-size relation is bellow the threshold for massive star formation. Meanwhile, core mass function (CMF) also exhibits a steeper slope towards high-mass end compared to more evolved core samples. These three properties in accordance suggest that although many IRDC cores are self-gravitating, only a small fraction are seemingly possible to form high-mass stars. In subsequent core evolution, some further mass assembly trend may be involved to facilitate the high-mass star formation. 

\end{abstract}

\keywords{\uat{
Interstellar medium}{847} --- \uat{Molecular clouds}{1072} --- \uat{Infrared dark clouds}{787} --- ISM: clouds, clumps and cores --- ISM: kinematics and dynamics}


\section{Introduction} 
\label{sect:intro}
Infrared dark clouds (IRDCs), seen as silhouettes against bright infrared emission, are ideal laboratories for studying the initial stages of massive star formation and the assembly of stellar clusters, including the environment in which low-mass stars form \citep{2018ARA&A..56...41M}.

We recently carried out the ALOHA IRDC survey (A Lei of the Habitat and Assembly of Infrared Dark Clouds), a systematic JCMT/SCUBA-2 dust-continuum survey of a carefully selected distance-limited sample of nearby IRDCs. As the first deep dust-mapping survey of this kind for IRDCs, ALOHA is designed to probe the initial conditions of Galactic massive star formation. The survey aims to reach the highest extended-source sensitivity achievable with SCUBA-2 ($\sim 2\,\mathrm{mJy\,beam^{-1}}$, close to the confusion limit), thereby mapping both the large-scale cloud environment and the internal dense structures of IRDCs. ALOHA provides unprecedented spatial resolution and intensity dynamic range, which are essential for recovering the full IRDC structure (dense cores and clumps), distinguishing evolutionary stages, measuring the intrinsic N-PDF, and constraining dust properties in massive star-forming regions. Each IRDC in this sample can be resolved down to the thermal Jeans scale.

The first key step is therefore to identify dense cores and clumps in an unbiased way and to measure their geometric and physical properties (e.g., size, mass, column density, and potential energy, together with their uncertainties). To achieve this goal, we developed the first version of the tool of Hierarchical cOre ideNtification and Kinematic property AssIgnment ({\sc honkai}). {\sc honkai} identifies structures either in PP (position--position) maps or in PPV (position--position--velocity) data cubes, and then derives physical properties for each identified structure. These properties can be calculated either from the same data product used for structure identification or from a user-specified auxiliary data set. In this work, we identify 2D structures in the ALOHA dust-continuum maps and use the MWISP $^{13}$CO cube as complementary 3D gas information to derive the gas properties associated with the dust structures.

The paper is organized as follows. In Sections \ref{aloha} and \ref{obs}, we introduce the ALOHA survey and the observations. In Section \ref{honkai}, we describe {\sc honkai} and the identification of structures in the ALOHA maps. In Sections \ref{mwisp}, \ref{xmatch}, and \ref{derive}, we use the MWISP $^{13}$CO data cube and cross-match it with the ALOHA dust data to derive the physical properties of the identified cores. In Section \ref{discussion}, we discuss the virial state of the cores and the core mass function (CMF). Our conclusions are summarized in Section \ref{summary}.

\section{ALOHA IRDCs}
\label{aloha}
The early phases of massive star formation ($M_*>8~M_\odot$) remain one of the central topics in astrophysics. IRDCs are widely regarded as the natal environments of massive stars and are therefore key targets for studying this problem. Extensive infrared observations of IRDCs have been obtained with facilities such as {\it ISO}, {\it MSX}, {\it Spitzer}, and {\it Herschel}. However, a critical missing ingredient is a systematic, deep, and high-angular-resolution survey at 850\,$\micron$, where the emission traces the Rayleigh--Jeans side of the spectral energy distribution (SED) and reveals the large-scale cloud structure with relatively low dust opacity.

IRDCs are believed to be the birth sites of stars over a broad range of masses \citep{2009ApJS..181..360C,2010ApJ...723L...7K, 2013A&A...552A..40C, 2014MNRAS.439.3275W}
, and many of them already show signposts of high-mass star formation \citep{2006A&A...447..929P,2002ApJ...566..945B}. Seen as silhouettes against the bright infrared background, IRDCs are considered to represent the initial conditions of massive star formation \citep{2006ApJ...641..389R, 2018ARA&A..56...41M}
Previous studies have also shown that a substantial fraction of the gas in IRDCs is concentrated in massive clumps, in some cases exceeding the empirical threshold for high-mass star formation \citep{2009A&A...505..405P}. IRDCs are therefore ideal targets for investigating the initial conditions of massive star formation, the evolutionary sequence of dense clouds, the aggregation of gas, and the assembly of dense cores.

We compiled a large sample of IRDCs with weak or undetected infrared emission as potential targets. We inspected their infrared images, in particular the {\it Spitzer}/IRAC 8\,$\micron$ absorption features together with the {\it Herschel} data \citep{2009A&A...505..405P}, and then selected a subsample of nearby, high-column-density clouds whose internal structures can be spatially resolved with JCMT/SCUBA-2. This refined sample is especially well suited for studying the initial conditions of massive star formation. We therefore carried out a deep 850\,$\micron$ SCUBA-2 survey toward these IRDCs. The target sensitivity of 2\,mJy\,beam$^{-1}$ is close to the deepest level achievable without severe confusion and is essential for imaging both extended structures and compact density peaks. It is also 5--10 times deeper than existing Galactic SCUBA-2 surveys such as the JCMT Plane Survey \citep{2017MNRAS.469.2163E} and SCUBA-2 Continuum Observations of Pre-protostellar Evolution \citep{2019MNRAS.485.2895E}. ALOHA is therefore designed to be the deepest large-area, spatially resolved 850\,$\micron$ continuum survey of a complete distance-limited sample of IRDCs in the early stages of potential massive star formation.

\section{Observation and Data Reduction}
\label{obs}
The observation of the ALOHA IRDCs is carried out with the JCMT SCUBA-2 bolometer at 850 $\micron$ band. Each observing block is a Pong 900 mapping, which covers a circular field with 20-arcmin diameter. The most observations were carried out in weather conditions better than band-3 ($\tau<0.12$). The integration time for each single field varies between $t_{\rm int}=3$ and 12 hours. We used the standard procedure in Starlink package\footnote{http://starlink.eao.hawaii.edu} to calibrate data and produce images. The noise level varies between 10 to 15 $\mjybm$ as due to the variation in $t_{\rm int}$ and weather conditions.

\begin{figure*}
	\centering
	\includegraphics[width=0.9\textwidth]{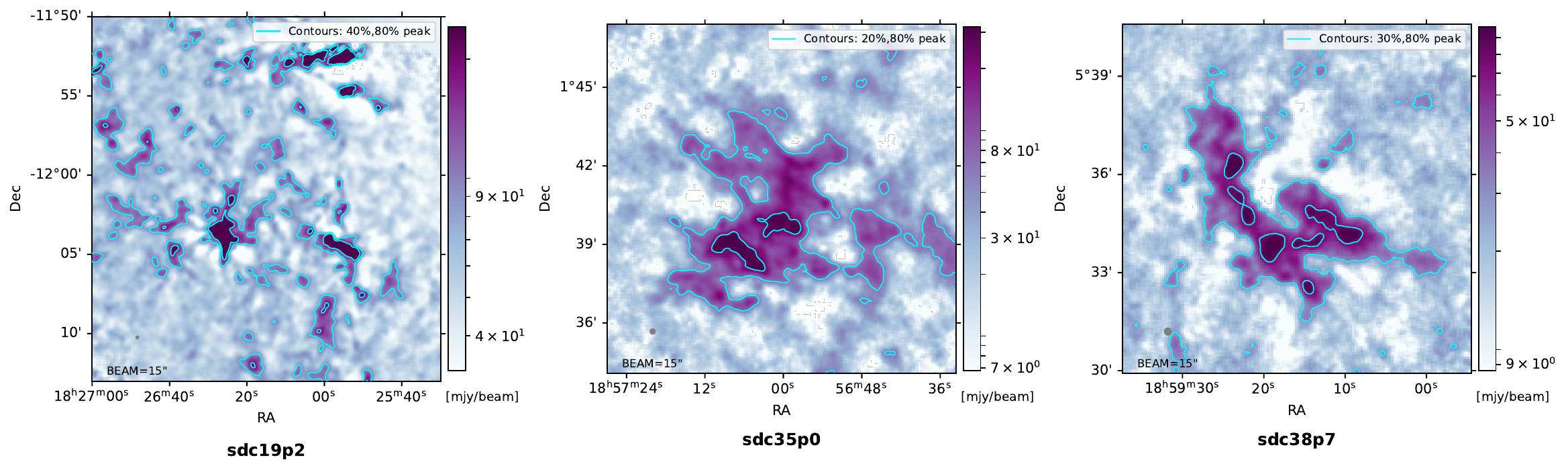}
	\caption{The maps display the intensity map of three representative star-forming regions from ALOHA observations: sdc19p2 (left), sdc35p0 (center), and sdc38p7 (right). Bright cyan contours outline specific peak-intensity levels. These panels clearly reveal both dense cores and clumps.}                                          
	\label{fig 1}
\end{figure*}

In this work, three representative star-forming regions from ALOHA observations are chosen to conduct analysis: sdc19p2, sdc35p0 and sdc38p7 (Fig \ref{fig 1}).

\section{IDENTIFICATION OF STRUCTURES IN ALOHA WITH honkai}
\label{honkai}
To identify cores and clumps in IRDCs in a reliable and automated way, we developed {\sc honkai} (tool of Hierarchical cOre ideNtification and Kinematic property AssIgnment) by combining the strengths of SCOUSEPY (Semi-automated multi-COmponent Universal Spectral-line fitting Engine) and ACORNS (Agglomerative Clustering for ORganising Nested Structures) \citep{2016MNRAS.457.2675H, 2019MNRAS.485.2457H}. SCOUSEPY is a semi-automated tool for fitting large quantities of complex spectroscopic data in a systematic and efficient manner, and for extracting Gaussian components from the spectrum at each pixel. ACORNS is a hierarchical $n$-dimensional clustering algorithm designed to organize decomposed spectroscopic data into nested structures. Full descriptions of SCOUSEPY and ACORNS are provided in Appendices \ref{Appendix A} and \ref{Appendix B}. These tools are particularly suitable for our purpose because they operate on decomposed spectroscopic data rather than directly on data cubes, allowing us to use not only intensity and position, but also centroid velocity, velocity dispersion, and other kinematic information in the structure-identification process. Because the method follows a bottom-up hierarchical agglomerative clustering (HAC) approach, it naturally captures multi-scale hierarchical structure, accurately extracts irregularly shaped features, and helps separate overlapping components. In addition, it retains pixel-scale information, making it possible to study how kinematic quantities vary across each identified structure \citep{2013A&A...554A..55H,2019MNRAS.485.2457H,2023ApJ...944L..15S}.

{\sc honkai} identifies structures using either PP (position--position) maps or PPV (position--position--velocity) data cubes, and then computes physical properties for each identified structure. The physical-property calculations can be carried out using the same data product that is used for structure identification, or using a user-specified auxiliary data set. In this paper, we first identify 2D structures on the high angular resolution ALOHA dust-continuum maps, and then connect them to independently identified 3D gas structures in the Milky Way Imaging Scroll Painting (MWISP) $^{13}$CO PPV cube to derive the gas properties associated with each dust structure.

\subsection{STRUCTURE IDENTIFICATION of ALOHA DUST DATA}
With the 15$''$ angular resolution of the ALOHA data, we are able to resolve dense cores and clumps within IRDCs. When analyzing PP data, {\sc honkai} performs clustering directly on the dust map. We set the minimum cluster radius (here and throughout this paper, ``cluster'' is used in the statistical sense to refer to an agglomeration of data points used to identify a molecular-cloud structure) to 18$''$, which is 20\% larger than the ALOHA beam. This ensures that all identified structures are spatially resolved. For two data points to be considered linked, we require their Euclidean separation to be no greater than 18$''$. This threshold is chosen to reflect the observational resolution.

The clusters identified in the initial phase are used to establish the first level of the hierarchy. We then relax the spatial linking criterion by 50\% to further develop the hierarchical structure.

\begin{figure*}
	\centering
	\includegraphics[width=0.9\textwidth]{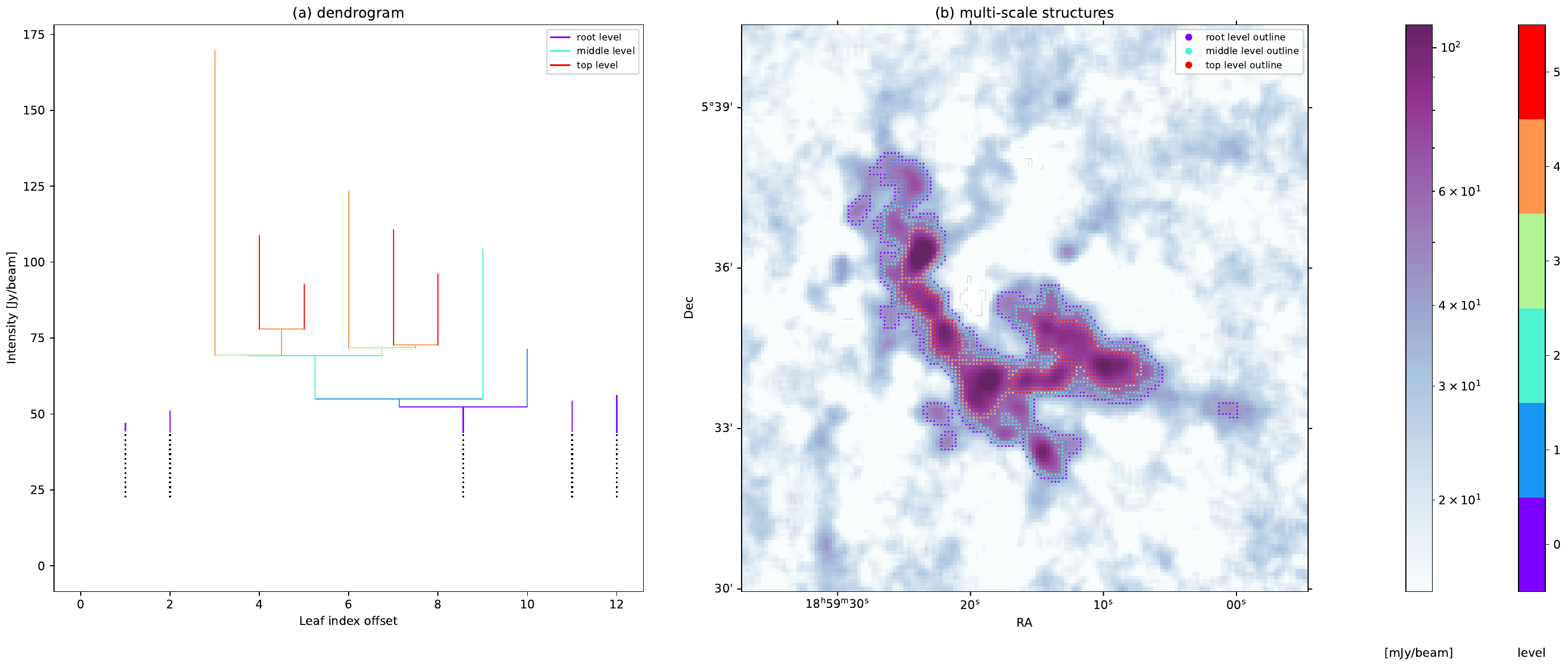}
	\caption{Visualization of the dust structure dendrogram hierarchy and its spatial footprint. (a) The ACORNS dendrogram built from the dust map, where branches are colored by hierarchical level. (b) The dust intensity map as the background, with multi-scale structure outlines overlaid in the same level-based color corresponding to the dendrogram. The colorbars indicate dust intensity and dendrogram level, respectively.}                                          
	\label{fig dendro}
\end{figure*}

As in other hierarchical clustering analyses, the result can be visualized as a dendrogram \citep{2008ApJ...679.1338R}. Figure \ref{fig dendro}(a) shows the ACORNS hierarchy derived from our data. Structures at different spatial scales are organized into a hierarchy system, in which compact dense cores are embedded within more diffuse large-scale clumps. Figure \ref{fig dendro}(b) shows the same structures projected into PP space. The identified trees closely follow the dominant features in the background dust-emission map. In the following analysis, we focus on two representative hierarchical levels: dense cores at the terminal branches of the hierarchy and larger clumps associated with the roots of the hierarchy.

\section{Physical properties of cores and clumps}
\label{mwisp}
\begin{figure*}
	\centering
	\includegraphics[width=0.9\textwidth]{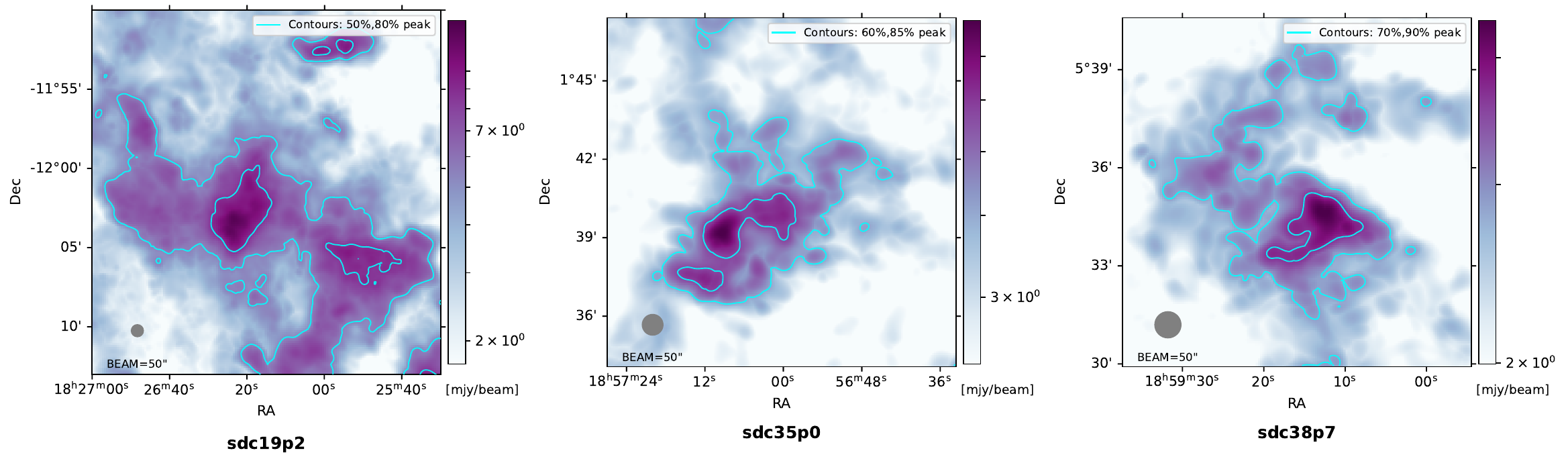}
	\caption{The maps display the peak intensity map of the same regions (as Fig \ref{fig 1}) from MWISP: sdc19p2 (left), sdc35p0 (center), and sdc38p7 (right). Bright cyan contours outline specific peak-intensity levels. These complex structures will be further decomposed with {\sc honkai}.}                                          
	\label{fig 2}
\end{figure*}

In this work, we use the $\rm ^{13}CO$ data cube from the Milky Way Imaging Scroll Painting (MWISP) survey \citep{2019ApJS..240....9S} to derive the physical properties of the cores and clumps identified in the ALOHA dust maps. MWISP is an unbiased Galactic-plane $\rm CO$ survey covering $l=-10^\circ$ to $+250^\circ$ and $|b|\lesssim 5^\circ.2$. The $\rm ^{13}CO$ data were obtained with the 13.7-m millimeter-wave telescope of the Purple Mountain Observatory (PMO), with an rms noise level of $\sim 0.3$\,K and a channel width of $\sim 0.17\,\mathrm{km\,s^{-1}}$. The beam size and pixel size are $\sim 50'' \times 50''$ and $\sim 30'' \times 30''$, respectively (Fig. \ref{fig 2}).

\subsection{Spectral decomposition of the MWISP data}
After the ALOHA cores and clumps have been identified, {\sc honkai} processes the MWISP $\rm ^{13}CO$ cube. In the first stage, the spectrum at each pixel is decomposed into Gaussian components. The procedure can be divided into several steps. {\sc honkai} first defines the spatial region over which fitting will be performed. This region can be specified by the user to target localized areas in both position and velocity, or to include only data above a chosen noise threshold in order to reduce the computational cost. {\sc honkai} then divides the map into small areas, referred to as spectral averaging areas (SAAs), and extracts an averaged spectrum from each one. The averaged spectrum of each SAA is fitted first, and the resulting solutions are supplied to the fully automated fitting routine that targets the individual spectra within that region. This process is controlled by a set of tolerance parameters. In this work, we adopt the following criteria: (1) all detected components must have a peak intensity greater than three times the noise level; (2) each Gaussian component must have a full-width at half maximum (FWHM) of at least one channel; and (3) two Gaussian components are considered distinguishable only if they are separated by at least half of the FWHM of the narrower component. 

\subsection{Decomposition of the MWISP structures}
In the next stage, {\sc honkai} further characterizes the cloud velocity structure using the spectral components extracted in the previous step. When clustering the PPV data, velocity information is included together with position. We set the minimum cluster radius to 60$''$, which is 20\% larger than the MWISP beam size, to ensure that all identified structures are spatially resolved. In addition to spatial proximity, we require consistency in velocity. Two data points are considered linked only if their Euclidean separation is no greater than 60$''$ and the absolute differences in both centroid velocity and velocity dispersion are no greater than $0.17\,\mathrm{km\,s^{-1}}$. These limits reflect the observational resolution of the data. Figure \ref{ppv} presents the identified structures in PPV space.

\section{Cross-matching 2D dust structures with 3D \texorpdfstring{$^{13}$CO}{13CO} structures}
\label{xmatch}
We have previously identified 2D structures (e.g., cores and clumps) on the ALOHA dust-continuum maps. However, a reliable derivation of their internal physical properties requires the additional kinematic and excitation information contained in 3D molecular-line data. We therefore employ the MWISP $^{13}$CO position-position-velocity (PPV) data cube. By applying Gaussian-component extraction and subsequent clustering, we independently identify coherent $^{13}$CO structures in PPV space, effectively resolving different velocity components that spatially overlap along the line of sight.

We then cross-match the 2D dust structures with the 3D gas structures identified in PPV space to derive the physical properties of each dust structure. For each dust structure, we search for all $^{13}$CO structures whose sky-projected footprints overlap with it. A key difficulty is that structures identified in a PPV cube can overlap when projected onto the sky, so multiple candidate $^{13}$CO components may intersect a single dust structure. To select the gas component that best corresponds to a given ALOHA dust structure, we use Spearman's rank correlation coefficient to quantify the morphological similarity between the dust continuum emission and the $^{13}$CO emission.

Our selection proceeds as follows. We first select candidate $^{13}$CO structures whose projected footprints cover at least $80\%$ of the dust-structure area. We then calculate the Spearman rank correlation coefficient for the overlapping pixels between the dust map and each candidate, and adopt the component with the highest correlation as the best match. If no candidate meets the 80\% coverage threshold, we relax the requirement to 50\% and again select the candidate with the highest Spearman correlation.

\begin{figure*}
	\centering
	\includegraphics[width=0.9\textwidth]{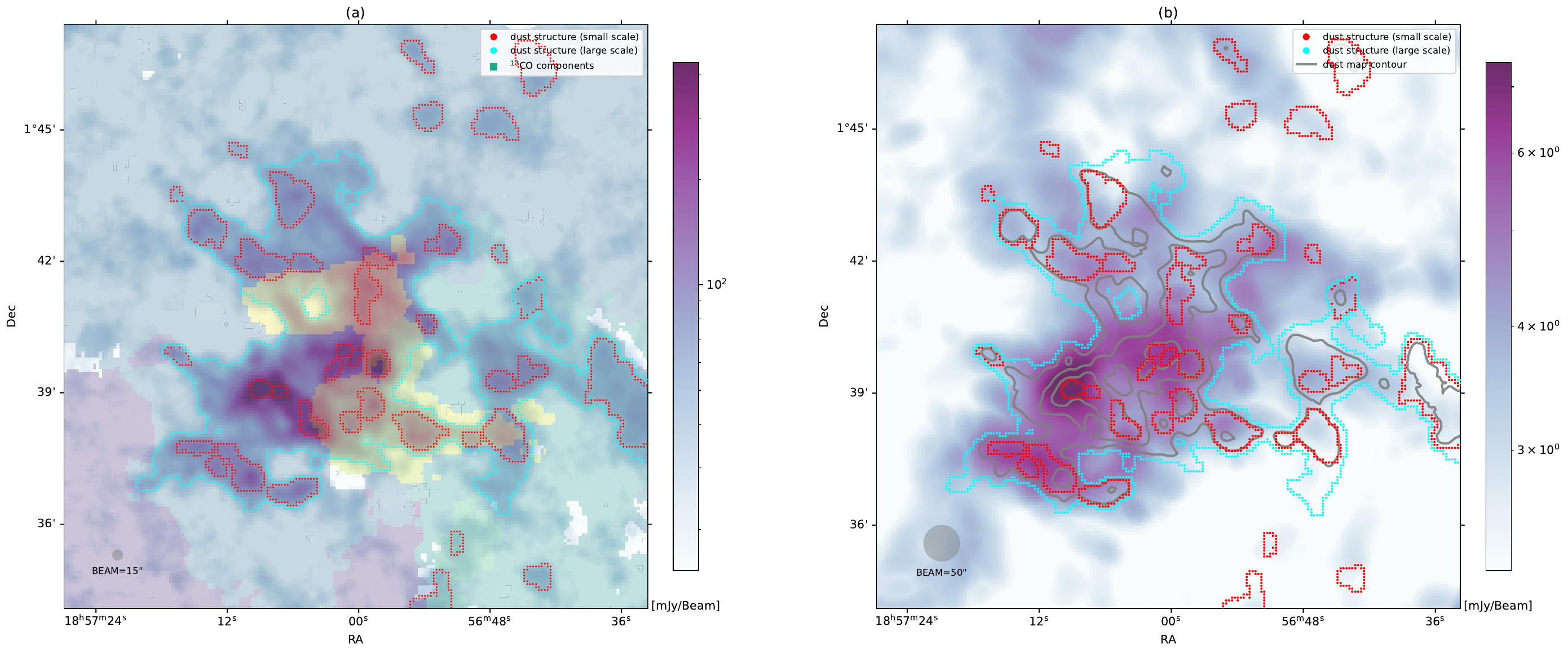}
	\caption{Identification results of ALOHA 2D dust structures and the MWISP 3D $^{13}$CO structures in sdc35p0. The red and cyan outlines display the boundaries of the identified small scale and large scale 2D dust structures. Left: The background image displays the ALOHA dust-continuum emission. The translucent colored patches represent the footprints of different $^{13}$CO velocity components identified in PPV space. Right: The background image shows the $^{13}$CO peak emission map. The gray contours trace the ALOHA dust-continuum emission.}                                          
	\label{fig 4}
\end{figure*}

\begin{figure*}
	\centering
	\includegraphics[width=0.9\textwidth]{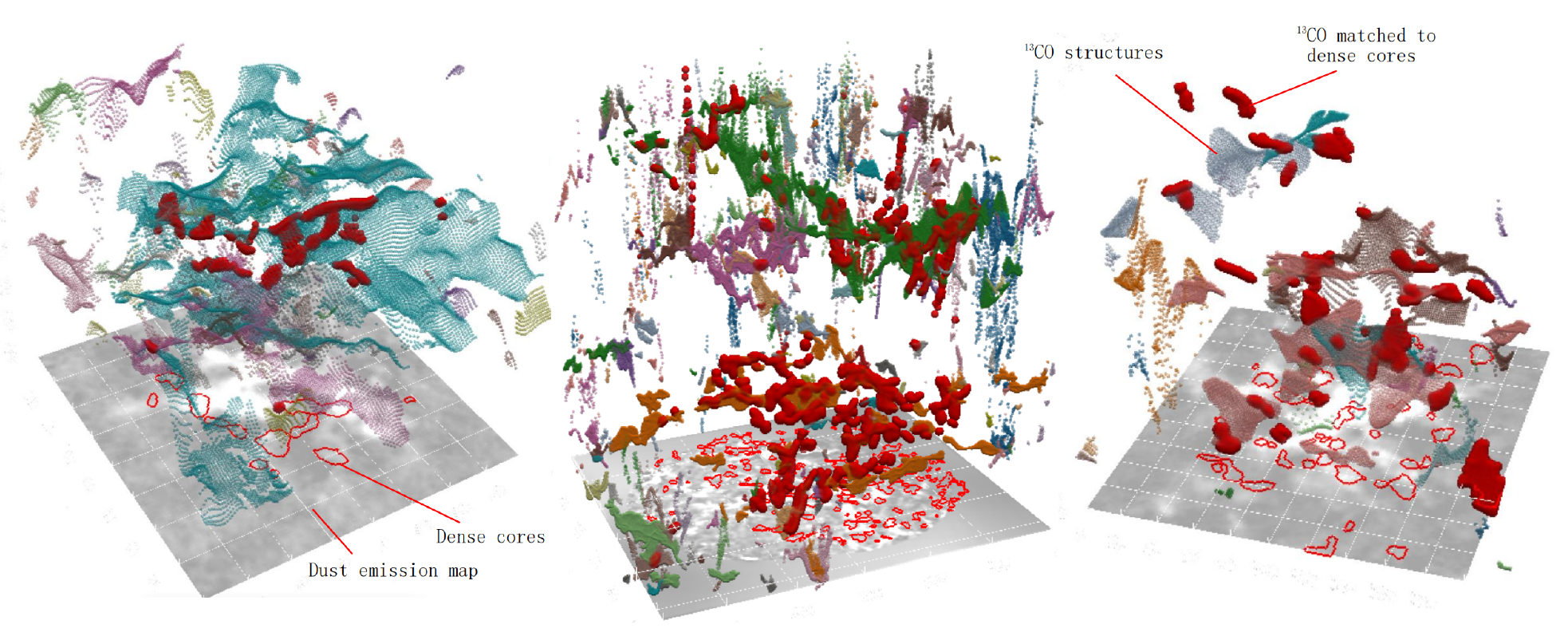}
	\caption{The cross-matching between 2D dust map and $^{13}$CO in PPV space. From left to right: sdc38p7, sdc19p2, sdc35p0. The background is the 2D dust emission map, dense cores are outlined in red. Different $^{13}$CO structures are marked with colors In PPV space, and the component matched to dense cores are highlighted with red.}                                          
	\label{ppv}
\end{figure*}

If no candidate reaches even 50\% coverage, we do not enforce a match for that dust structure. Instead, for an approximate estimate of the subsequent physical properties, we directly use the intensity and line width from the SCOUSEPY Gaussian decomposition of the $^{13}$CO data within the spatial extent of the dust structure. When multiple Gaussian components are fitted in a single pixel, we adopt the component with the maximum intensity. Only 5 out of 193 dust structures used this rough estimation.
An example of the identified structures are shown in Figs. \ref{fig 4}, and an overview of the cross-matching in PPV space is presented in Fig. \ref{ppv}.


\section{Derivation of Physical parameters}
\label{derive}
Once the correspondence has been established, we extract from the matched $^{13}$CO structure the Gaussian components on a pixel-by-pixel basis within the overlap region. The matched pixel-level Gaussian components are then used in the subsequent physical-property calculations.


(1) The column density is derived from the dust continuum emission under the assumption of optically thin emission:
\begin{equation}
N(\mathrm{H}_2) = \frac{I_\nu}{\kappa_\nu \, B_\nu(T_d) \, \mu_{\mathrm{H}_2} \, m_{\mathrm{H}}}
\end{equation}
where $I_\nu$ is the observed intensity, $\kappa_\nu$ is the dust opacity at 850\,$\mu$m, $B_\nu(T_d)$ is the Planck function at dust temperature $T_d$, $\mu_{\mathrm{H}_2}=2.8$ is the mean molecular weight per hydrogen molecule, and $m_{\mathrm{H}}$ is the mass of a hydrogen atom. In this work, we adopt $\kappa_{850}=0.015\,\mathrm{cm^2\,g^{-1}}$. The dust temperature $T_d$ is taken from Ren et al. (in prep.), and in general it can be estimated from multi-wavelength spectral energy distribution (SED) fitting.

(2) The volume number density is estimated as $n_0 = N(\mathrm{H}_2)/(2r_{\rm eff})$.   


(3) The total mass is calculated from the dust continuum emission by summing over all pixels within each structure:
\begin{equation}
M = \sum \frac{S_{\nu,\mathrm{pix}} \, D^2}{\kappa_\nu \, B_\nu(T_{d,\mathrm{pix}})} ,
\end{equation}
where $S_{\nu,\mathrm{pix}}$ is the flux density in each pixel. The Kinematic distances $D$ were derived for all IRDCs with the BeSSeL Bayesian distance estimator, using the source coordinates and $^{13}$CO $V_{LSR}$ as inputs. The estimator is based on the parallax-constrained Milky Way model of \cite{2019ApJ...885..131R} and incorporates information on Galactic rotation and spiral-arm structure. For the present sample, we adopted the near-distance solution, as IRDCs seen in mid-infrared absorption are usually located on the near side of the Galaxy, where they can effectively obscure the diffuse Galactic background emission.


\subsection{Dynamic State of cores}
An important question is whether the identified cores are gravitationally bound. To address this, we calculate their gravitational potential energy. For an axisymmetric ellipsoid of mass $M$ with a concentric density profile, the potential energy is 
\begin{equation}
G = -\frac{3}{5} \alpha \beta \frac{GM^2}{r}
\end{equation}
where $r$ is the radius, $\beta=\arcsin e/e$ is the geometrical factor determined by the eccentricity $e=\sqrt{1-f^2}$, $\alpha=(1-a/3)/(1-2a/5)$ for a power-law density profile $\rho\propto r^{-a}$, and $G$ is the gravitational constant. We adopt $a=1.6$ for an isothermal cloud in equilibrium following \cite{1956MNRAS.116..351B}. Because of projection effects, the intrinsic axis ratio $f$ is generally smaller than the observed ratio $f_{\text{obs}}$. We therefore use
\begin{equation}
f = \frac{2}{\pi} f_{\text{obs}} \mathcal{F}_1(0.5, 0.5, -0.5, 1.5, 1, 1 - f_{\text{obs}}^2)
\label{eq:axis_ratio}
\end{equation}
which gives the most likely value for a prolate ellipsoid \citep{1983ApJ...271..417F}, where $\mathcal{F}_1$ is the Appell hypergeometric function of the first kind.

Because the identified cores have irregular morphologies, we derive an equivalent eccentricity from an intensity-weighted image-moment analysis. For each core, we construct a two-dimensional spatial covariance matrix from the pixel coordinates weighted by the local intensities. The equivalent major and minor axes are then obtained from the square roots of the eigenvalues ($\lambda_1$ and $\lambda_2$) of this matrix. This approach provides a robust estimate of the core eccentricity.

For cores with negligible magnetic support and no external pressure confinement, the virial mass $M_{\rm vir}$ for a given velocity dispersion $\sigma$ and radius $r$ is
\begin{equation}
M_{\text{vir}} = \frac{5}{\alpha \beta} \frac{\sigma^2 r}{G}
\end{equation}
Cores with $M>M_{\rm vir}$ are too massive to remain in virial equilibrium. We quantify this balance using the virial mass ratio
\begin{equation}
R_{\text{vir}} = \frac{M}{M_{\text{vir}}}
\end{equation}

The Jeans mass provides an approximate upper limit for the stability of a gas clump or core. Above this limit, self-gravity overcomes internal pressure support and even a small perturbation may trigger collapse or fragmentation. For a non-magnetic isothermal cloud, it can be estimated as \citep{1990sse..book.....K}
\begin{equation}
M_{J} = \frac{\pi ^{5/2} c_{\rm s,eff}^3}{6G^{3/2}\rho^{1/2}}
\end{equation}
where $r_{\rm eff}=\sqrt{A/\pi}$ is the effective radius, $\rho=\mu m_{\rm H} n_0$ is the average density, and $c_{\rm s,eff}$ is the effective sound speed including both thermal and non-thermal gas motions. The thermal and non-thermal components are $\sigma_{\rm th}=(kT_{\rm kin}/\mu m_{\rm H})^{1/2}$ and $\sigma_{\rm nth}=(\sigma_{\rm obs}^2-kT_{\rm kin}/m_{\rm mol})^{1/2}$, respectively. 

\section{Discussion}
\label{discussion}
\begin{figure*}
	\centering
	\includegraphics[width=0.9\textwidth]{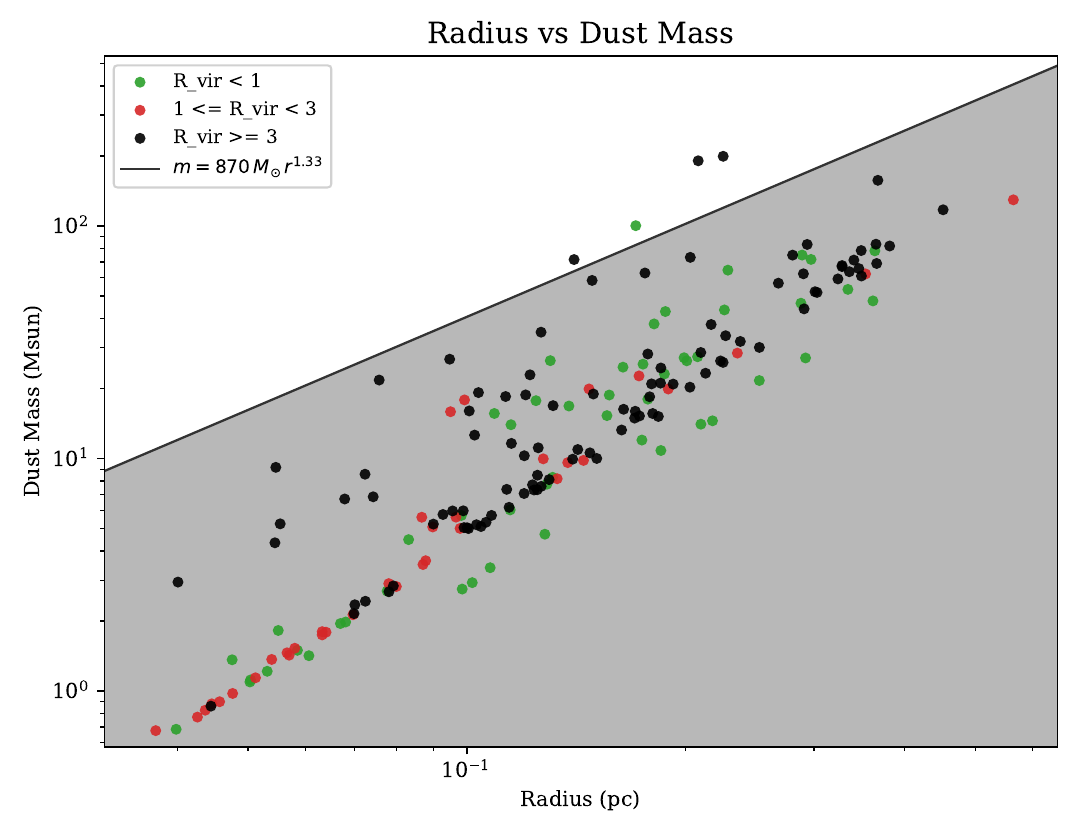}
	\caption{The core mass-size relationship. The colors denote the virial mass ratio $R_{\text{vir}}$, with green for $R_{\text{vir}}<1$, red for $1<R_{\text{vir}}<3$, and black for $R_{\text{vir}}>3$. The solid curve represents the empirical threshold for massive star formation \citep{2010ApJ...723L...7K}. }                               
	\label{Mvir}
\end{figure*}

We find that 136 of the 193 cores (70\%; Table \ref{tab:core_catalog}) have $R_{\text{vir}}>1$, and 101 cores (52\%) have $R_{\text{vir}}>3$. The three most extreme cases have $R_{\text{vir}}>45$. We further compare our sample with the empirical threshold for massive star formation proposed by \cite{2010ApJ...723L...7K} in Fig. \ref{Mvir}. In our sample, the core radii are generally distributed between 0.1 and 1\,pc, and only four objects exceed the empirical threshold. This trend is consistent with the statistical result presented in Fig.~4 of Kauffmann \& Pillai (2010), where most cores in a similar size range also remain below the threshold for massive star formation. These results suggest that although most cores in our IRDC sample are gravitationally bound and likely unstable against collapse, only a small fraction are plausible candidates for massive star formation.

\begin{figure*}
	\centering
	\includegraphics[width=0.9\textwidth]{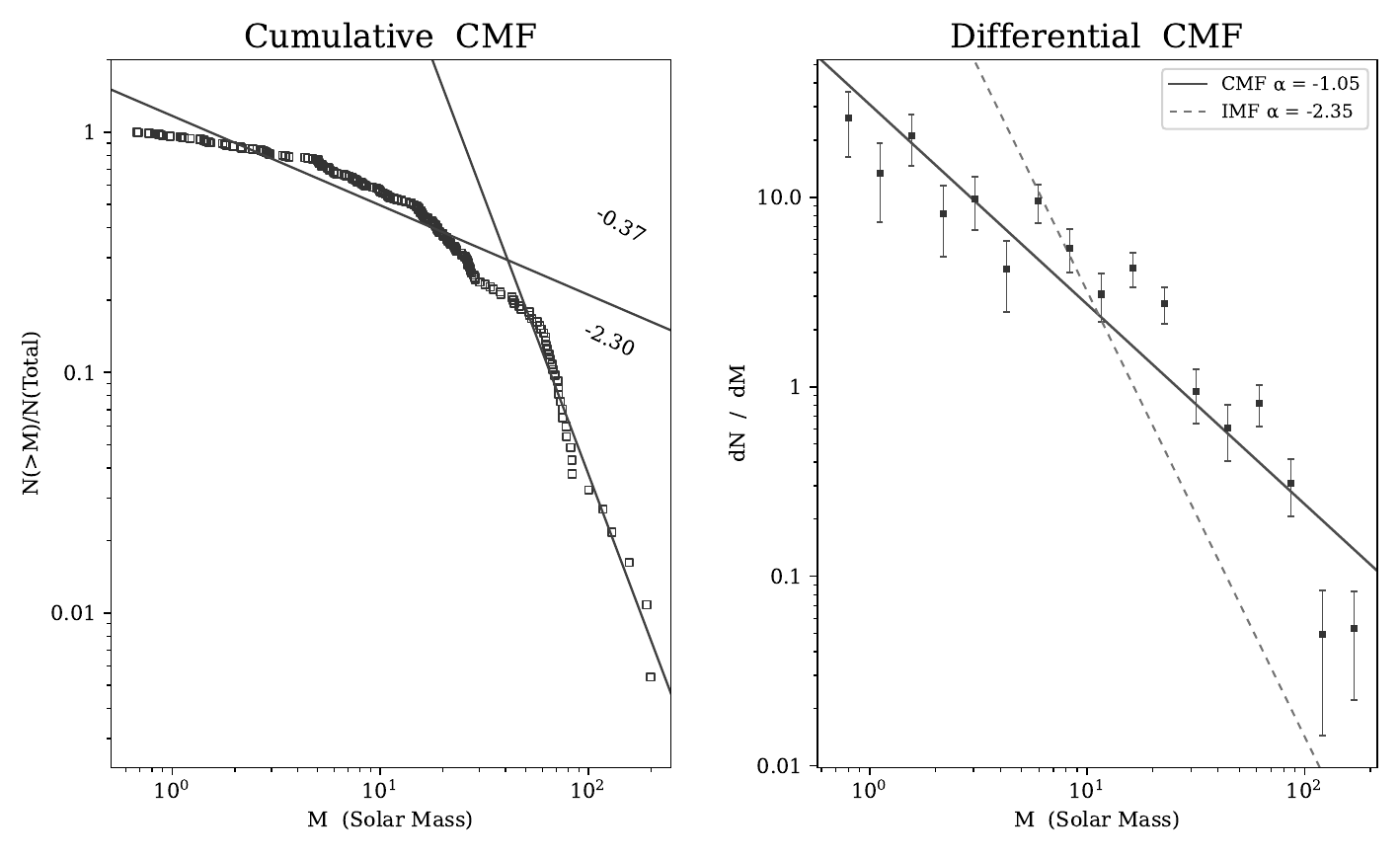}
	\caption{Left: Cumulative Core Mass Function with the two power laws fitted to different parts of the data. The power-law index for each fitted line is labeled in the plot. Right: Differential Core Mass Function. The dashed line shows the $-2.35$ slope of a Salpeter stellar IMF, and the solid line is the best fit to the binned differential CMF with a slope of $-1.1$. }           
	\label{cmf}
\end{figure*}

The core mass function (CMF) describes the statistical distribution of dense-core masses and is commonly expressed as $dN/d\log M$ (or $dN/dM$). It provides an important observational constraint on cloud fragmentation and on the origin of stellar masses. With {\sc honkai}, we compute core masses for all identified structures in a uniform way, construct the CMF for the current ALOHA sample, and fit its functional form to characterize the mass scale and slope. In Fig. \ref{cmf}, we present the CMF in two forms: (1) the cumulative mass function, which measures the fraction of cores with masses above a given value, and (2) the differential mass function, in which the data are binned into specified mass intervals. The cumulative CMF suggests a broken power-law behavior, with a steep slope of $-3.4$ at the high-mass end and a flatter slope of $-1.4$ at lower masses. The high-mass slope is steeper than the value of about $-2.2$ reported in several previous studies \citep{2005ApJ...625..891R, 2007ApJ...655..351L}. By contrast, a single power-law fit to the differential CMF yields a slope close to $-1.1$. 

By combining the JCMT 850\,$\micron$ dust continuum with the MWISP isotopologue CO $(1-0)$ spectral-line data, we demonstrate a synchronized identification of very early cold cores together with homogeneous measurements of their kinematic properties. Compared with earlier studies that relied primarily on a single tracer \citep{2019ApJ...886..102S}, the present dust--molecular-line joint analysis provides significantly stronger statistical constraints on the initial dynamical conditions. In particular, we find that self-gravity and turbulent energy are comparable for most of the identified cores, supporting the view that molecular clouds can already be subject to substantial self-gravitational confinement at a very early evolutionary stage. However, only a few cores lie above the critical mass--radius relation associated with high-mass star formation, and the CMF of this sample is steeper than those reported for massive cores in the hot-core and H\textsc{ii}-region stages. Taken together, these results suggest that IRDCs may not, in most cases, form high-mass stars directly from their current core population, but instead depend more strongly on subsequent mass assembly and inward material transport during later evolution.

The modular design of {\sc honkai} also makes it well suited for future large observational programs. As increasingly sensitive continuum and spectral-line surveys become available, {\sc honkai} can be applied to larger dust-continuum samples and can be extended to other tracers that probe complementary density and temperature regimes. In this sense, {\sc honkai} can provide a practical tool for the scientific exploitation of next-generation facilities such as {\it JWST} and {\it CSST}, where structure identification, cross-matching between imaging and spectroscopic data, and automated physical-property measurements will become even more important for handling large and diverse data sets.

We note that the present IRDC sample covers only three regions. A full ALOHA catalog and a more comprehensive statistical analysis will be presented in future work. The catalog of cores and their properties in our sample can be found in Appendix \ref{catalog}.

\section{Summary}\label{summary}
In this work, we present {\sc honkai}, an automatic measurement pipeline designed to identify structures in PP and PPV data and to derive their physical properties in a homogeneous way. Applying this method to the high-sensitivity 850\,$\micron$ ALOHA dust-continuum data and the MWISP $^{13}$CO cube, we carried out a joint analysis toward three representative IRDC regions, namely sdc19p2, sdc35p0, and sdc38p7.

Our main results can be summarized as follows. (1) With the combination of hierarchical clustering and multi-wavelength cross-identification, {\sc honkai} is able to extract cores and clumps over a wide range of spatial scales and to connect them with associated gas components in PPV space. (2) Based on the matched dust and gas information, we present a catalog of 193 dense cores and 16 clumps with positions, sizes, kinematics, masses, densities, virial parameters, Jeans masses, and morphologies. (3) The joint JCMT 850\,$\micron$ and MWISP isotopologue CO $(1-0)$ analysis enables synchronized identification of very early cold cores and homogeneous measurements of their kinematic properties, providing stronger statistical constraints on the initial dynamical conditions than single-tracer studies \citep{2019ApJ...886..102S}. (4) The virial analysis shows that a large fraction of the identified cores are gravitationally bound and likely unstable to collapse: 136 of 193 cores have $R_{\rm vir}>1$, and 101 cores have $R_{\rm vir}>3$. Self-gravity and turbulent energy are comparable for most cores, but only four cores lie above the empirical mass--size threshold for massive star formation. (5) The CMF derived from the current sample shows a cumulative high-mass slope steeper than several previous measurements, suggesting that IRDC cores may not directly form high-mass stars in most cases and may instead require subsequent mass assembly and inward material transport during later evolution; the differential CMF is characterized by a power-law index close to $-1.1$.

These results highlight {\sc honkai}'s impact as an efficient and reproducible framework for identifying cores and clumps and for measuring their physical properties with multi-wavelength data. With the future release of the full ALOHA sample, the method presented here will enable a more statistically robust characterization of the dense-core population, their dynamical states, and the origin of the CMF in the early phases of massive star formation. More generally, as more sensitive continuum and spectral-line data become available, {\sc honkai} can be extended to larger samples and other tracers, and may also support future studies with facilities such as {\it JWST} and {\it CSST}.

\begin{acknowledgments}
Supported by NSFC 125888202. DL acknowledges support from the New Cornerstone foundation. 

\end{acknowledgments}

\begin{contribution}

All authors contributed equally to the collaboration.


\end{contribution}

%
\facilities{JCMT, PMO}

\software{astropy \citep{2013A&A...558A..33A,2018AJ....156..123A,2022ApJ...935..167A},  
          Scousepy \citep{2016MNRAS.457.2675H}, 
          Acorns \citep{2019MNRAS.485.2457H}
          }


\appendix

\section{Appendix A}
\label{Appendix A}
The SCOUSEPY procedure is described in detail by \cite{2016MNRAS.457.2675H}; here we summarize only the main steps relevant to this work. Briefly, SCOUSEPY first defines the spatial region over which fitting will be performed. This region can be tailored by the user to target localized areas in both position and velocity, or to include only data above a specified noise threshold, thereby reducing the fitting workload. SCOUSEPY then divides the map into small regions, referred to as spectral averaging areas (SAAs), and extracts an averaged spectrum from each one. The user can refine the SAA size according to the local complexity of the line profiles, so that more complex regions are analyzed with smaller SAAs. This refinement improves the overall fitting quality, especially for large and complex data sets, because the initial guesses supplied to the automated fitting stage become more accurate. The spatially averaged spectra extracted from each SAA are then fitted interactively with {\sc PYSPECKIT}, whose extensible framework supports a variety of line-profile models, including Gaussian, Voigt, Lorentzian, and hyperfine-structure fitting. For the MWISP data used here, we assume that the spectra can be decomposed into Gaussian components. The best-fitting solutions obtained for the SAAs are then passed to the fully automated fitting procedure that targets all individual spectra within each region. This process is controlled by several tolerance levels. For a full description of these tolerances, see \cite{2016MNRAS.457.2675H}. In our analysis, we adopt the following criteria: (i) all detected components must have a flux density greater than three times the local noise value ($T_1=3.0$); (ii) each Gaussian component must have an FWHM of at least one channel ($T_2=1.0$); and (iii) two Gaussian components are considered distinguishable only if they are separated by at least half of the FWHM of the narrower component ($T_5=0.5$). The remaining two tolerance levels ($T_3$ and $T_4$) restrict the extent to which the parameters of individual velocity components may deviate from their closest counterparts in the SAA spectrum; we set both to 3.0. The final best-fitting solution for each pixel is chosen as the one with the minimum corrected Akaike information criterion \citep{1974ITAC...19..716A}. 

\section{Appendix B}
\label{Appendix B}
\subsection{Introduction and description of the input parameters}

A key feature of {\sc ACORNS} is that it operates on decomposed spectroscopic data rather than directly on data cubes. This makes it especially suitable for obtaining a detailed description of the gas kinematics.

The ACORNS procedure is described in detail by \cite{2019MNRAS.485.2457H}. ACORNS follows the philosophy of hierarchical agglomerative clustering (HAC). HAC methods are generally divided into two categories: bottom-up and top-down. ACORNS adopts the bottom-up approach, in which each individual data point initially forms a separate cluster and clusters are then merged to build a hierarchy.

Clustering in {\sc ACORNS} commences with the most significant data point. In this work, this refers to the data point with the greatest peak intensity, but for other systems it may instead refer to a density, column density, or mass. {\sc ACORNS} then descends in significance, merging clusters based on physically motivated, user-provided criteria until a hierarchy is established.

Input to {\sc ACORNS} is an array of $n \times m$ dimensions, where $n$ is the number of parameters (at minimum 4, but in principle with no upper limit) and $m$ is the number of data points. In its simplest form for clustering in two spatial dimensions, the array should consist of $x$ position, $y$ position, intensity (or equivalent), and an uncertainty on the intensity (or equivalent). If linking in PPV, an additional column for the velocity is required.

The linking of clusters is handled via an array containing $n-3$ elements (or $n-4$ if linking in PPP) that describes the clustering criteria. The user supplies the maximum spatial Euclidean distance between data points, as well as the maximum absolute difference in any other variable used for linking. If the separation between two data points satisfies these criteria, the data points are considered linked. No two data points extracted from the same location can be linked to the same cluster.

In addition, the user supplies the following parameters: (i) the pixel size in units consistent with the positional information in the input array; (ii) the radius of the smallest structures the user would like {\sc ACORNS} to identify; (iii) the minimum height above a merge level for a cluster to be considered a separate structure; and (iv) the stopping criterion, given as a multiple of the rms noise level $\mathrm{stop}$.

\subsubsection{A description of the method}
The main steps taken by {\sc ACORNS} in developing the hierarchy are as follows:
(1) {\sc ACORNS} creates a catalogue of currently unassigned data. All data whose intensity $I$ satisfies $I > \mathrm{stop} \times \sigma_{\mathrm{rms}}$ are added to this catalogue, where $\sigma_{\mathrm{rms}}$ is the noise level at that position. The unassigned data are then rearranged in descending order of $I$.
(2) These data are used to generate a k-d tree that can be queried to return the nearest neighbours to a given point.
(3) Starting with the first data point in the unassigned catalogue and looping over all data points, {\sc ACORNS} generates a 'bud cluster', it refers to a structure that has not yet met the criteria to become a 'leaf' in its own right (where leaves are the clusters situated at the top of the hierarchical system). It then queries the k-d tree to find all data points within the maximum Euclidean distance (provided in $\mathrm{cluster_criteria}$) from the bud cluster and checks any additional linking criteria supplied by the user. All data points satisfying the criteria are cross-referenced against the current cluster catalogue to see if they belong to an already established cluster; if so, a link is established and the hierarchy grows.
(4) Once {\sc ACORNS} has cycled through all data points in the unassigned catalogue, it begins a second loop. The cluster catalogue is cleaned of any bud clusters and these data are used to generate a new unassigned catalogue. This step picks up data points that were unable to be linked during the first pass.
(5) If specified by the user, the clustering criteria are relaxed and {\sc ACORNS} performs additional loops based on the new criteria. This helps further develop the hierarchy.
(6) {\sc ACORNS} discards all remaining bud clusters since they did not meet the criteria to become fully fledged clusters.

{\sc ACORNS} returns a system of clusters as its output. In a given hierarchy, the antecessor is the largest common ancestor of all clusters within that hierarchy. For a given data set there may be multiple antecessors, and each may or may not have descendant substructure. Expanding the nomenclature typically used to describe dendrograms, an antecessor refers to a tree in a forest of clusters. Each tree may or may not exhibit substructure, referred to as branches and leaves.

\subsubsection{The growth of the hierarchy}
After establishing a link between a bud cluster and already established clusters in the hierarchy, the next step depends on the number of linked clusters:
(1) If no linked clusters are identified, the bud cluster is added to the cluster catalogue as a new cluster.
(2) If a single linked cluster is identified, the bud cluster is merged into this already established cluster.
(3) If multiple linked clusters are identified, further decisions are required. {\sc ACORNS} first determines how many of the linked clusters are true clusters (not bud clusters). What happens next depends on how many fully fledged clusters the bud cluster is linked to: (i) if none, then all linked clusters are bud clusters and the bud cluster merges into the first of the other buds; (ii) if there is a single linked cluster, the bud cluster merges into that cluster; (iii) if multiple linked clusters are present, a branch is generated between these clusters to create a new level in the hierarchy. Any remaining bud clusters are then merged with the same cluster as the original bud cluster.

\subsubsection{Relaxing the linking constraints}
This optional second phase is important when working with discrete and irregularly spaced data, such as velocity information output following spectral decomposition. The idea is to relax the linking constraints used during the first stage in order to further develop the hierarchy. The user can relax the constraints either in a single step or incrementally.

{\sc ACORNS} first generates a new catalogue of data points that were not assigned to clusters during the first and second passes using the initial criteria. As in the earlier stages, {\sc ACORNS} starts with the most significant data point in the unassigned catalogue, and the same steps are implemented using the relaxed criteria. During this phase, {\sc ACORNS} attempts to link bud clusters to an already established forest, but it ensures that links remain strong despite the relaxed criteria. If additional properties are considered when searching for links (e.g., centroid velocity or velocity dispersion), {\sc ACORNS} checks these properties against those of the linked clusters. If the properties of the bud cluster lie more than $3\sigma$ away from the mean of the linked cluster properties, those links are prevented from forming.

During the relaxation phase, a bud cluster may be linked to multiple trees within the forest, or to multiple clusters within the same tree. {\sc ACORNS} then determines whether the bud cluster can be inserted at the correct position in the hierarchy, which is governed by the peak intensity of the linked clusters. If the bud cluster cannot be inserted directly into a linked cluster, {\sc ACORNS} searches downward through the hierarchical tree, where possible, to establish a valid link. If the bud cluster still cannot be placed at the correct level in any existing tree, those linked clusters are ignored. ACORNS then returns a single linked cluster per tree to which the bud cluster will be attached. If a new branch must be created, the bud cluster is first merged with the closest matching cluster among all available linked clusters, after which a new branch is created at the base of the parent hierarchies. 

\section{Catalog}
\label{catalog}
\begingroup
\movetabledown=0.4in
\begin{longrotatetable}
\startlongtable
\begin{deluxetable}{lllllllllllllll}
\tabletypesize{\tiny}
\tablewidth{0pt}
\tablecaption{Core catalog.\label{tab:core_catalog}}
\tablehead{
\colhead{ID} &
\colhead{$D$} &
\colhead{R.A.} &
\colhead{Dec.} &
\colhead{$V$} &
\colhead{$\Delta V$} &
\colhead{$c_{\rm eff}$} &
\colhead{$R_{\rm eff}$} &
\colhead{$\rho$} &
\colhead{$M$} &
\colhead{$M_{\rm vir}$} &
\colhead{$R_{\rm vir}$} &
\colhead{$M_{\rm J}$} &
\colhead{$e$} &
\colhead{$E_{\rm pot}$} \\
\colhead{} &
\colhead{(kpc)} &
\colhead{(J2000)} &
\colhead{(J2000)} &
\colhead{(km s$^{-1}$)} &
\colhead{(km s$^{-1}$)} &
\colhead{(km s$^{-1}$)} &
\colhead{(pc)} &
\colhead{(g cm$^{-3}$)} &
\colhead{($M_\odot$)} &
\colhead{($M_\odot$)} &
\colhead{} &
\colhead{($M_\odot$)} &
\colhead{} &
\colhead{(erg)}
}
\startdata
0       & 1.847                                             & 18:59:01                                               & 5:33:22                                                & 32.6                                               & 1.98                                                   & 0.87                                                     & 0.061                                                 & 1.0E-19                                                                 & 1.4                                                & 30.9                                                    & 0.05   & 174                                                       & 0.81 & -2.75E+42                                                    \\
1       & 1.676                                             & 18:59:22                                               & 5:32:46                                                & 29.7                                               & 0.66                                                   & 0.36                                                     & 0.055                                                 & 1.8E-19                                                                 & 1.8                                                & 3.1                                                     & 0.58   & 9.1                                                       & 0.74 & -4.90E+42                                                    \\
\enddata
\end{deluxetable}
\end{longrotatetable}

\begin{longrotatetable}
\startlongtable
\begin{deluxetable}{lllllllllllllll}
\tabletypesize{\tiny}
\tablewidth{0pt}
\tablecaption{Clump catalog.\label{tab:large_scale_catalog}}
\tablehead{
\colhead{ID} &
\colhead{$D$} &
\colhead{R.A.} &
\colhead{Dec.} &
\colhead{$V$} &
\colhead{$\Delta V$} &
\colhead{$c_{\rm eff}$} &
\colhead{$R_{\rm eff}$} &
\colhead{$\rho$} &
\colhead{$M$} &
\colhead{$M_{\rm vir}$} &
\colhead{$R_{\rm vir}$} &
\colhead{$M_{\rm J}$} &
\colhead{$e$} &
\colhead{$E_{\rm pot}$} \\
\colhead{} &
\colhead{(kpc)} &
\colhead{(J2000)} &
\colhead{(J2000)} &
\colhead{(km s$^{-1}$)} &
\colhead{(km s$^{-1}$)} &
\colhead{(km s$^{-1}$)} &
\colhead{(pc)} &
\colhead{(g cm$^{-3}$)} &
\colhead{($M_\odot$)} &
\colhead{($M_\odot$)} &
\colhead{} &
\colhead{($M_\odot$)} &
\colhead{} &
\colhead{(erg)}
}
\startdata
0       & 1.5                                               & 18:27:02                                               & -11:53:34                                              & 12.8                                               & 0.42                                                   & 0.28                                                     & 0.671                                                 & 1.7E-20                                                                 & 322.4                                              & 15.9                                                    & 20.27672956 & 14.7                                                      & 0.62 & -1.23E+46                                                    \\
1       & 1.5                                               & 18:25:51                                               & -12:14:45                                              & 12.7                                               & 0.27                                                   & 0.25                                                     & 0.455                                                 & 1.6E-20                                                                 & 232.1                                              & 4.6                                                     & 50.45652174 & 10.3                                                      & 0.48 & -6.82E+45                                                    \\
\enddata
\tablecomments{$V$: rest velocity; $\Delta V$: velocity dispersion; $c_{\rm eff}$: effective sound speed; $R_{\rm eff}$: effective radius; $\rho$: volumn density; $M$: mass; $M_{\rm vir}$: virial mass; $R_{\rm vir}=M/M_{\rm vir}$: virial mass ratio; $M_{\rm J}$: Jeans mass; $e$: eccentricity; $E_{\rm pot}$: gravitational potential energy.}
\end{deluxetable}
\end{longrotatetable}
\endgroup

\bibliography{HONKAI}{}

@ARTICLE{2018ARA&A..56...41M,
       author = {{Motte}, Fr{\'e}d{\'e}rique and {Bontemps}, Sylvain and {Louvet}, Fabien},
        title = "{High-Mass Star and Massive Cluster Formation in the Milky Way}",
      journal = {\araa},
         year = 2018,
        month = sep,
       volume = {56},
        pages = {41-82},
          doi = {10.1146/annurev-astro-091916-055235},
archivePrefix = {arXiv},
       eprint = {1706.00118},
 primaryClass = {astro-ph.GA},
       adsurl = {https://ui.adsabs.harvard.edu/abs/2018ARA&A..56...41M}
}

@ARTICLE{2009ApJS..181..360C,
       author = {{Chambers}, E.~T. and {Jackson}, J.~M. and {Rathborne}, J.~M. and {Simon}, R.},
        title = "{Star Formation Activity of Cores within Infrared Dark Clouds}",
      journal = {\apjs},
         year = 2009,
        month = apr,
       volume = {181},
       number = {2},
        pages = {360-390},
          doi = {10.1088/0067-0049/181/2/360},
       adsurl = {https://ui.adsabs.harvard.edu/abs/2009ApJS..181..360C}
}

@ARTICLE{2010ApJ...723L...7K,
       author = {{Kauffmann}, Jens and {Pillai}, Thushara},
        title = "{How Many Infrared Dark Clouds Can form Massive Stars and Clusters?}",
      journal = {\apjl},
         year = 2010,
        month = nov,
       volume = {723},
       number = {1},
        pages = {L7-L12},
          doi = {10.1088/2041-8205/723/1/L7},
archivePrefix = {arXiv},
       eprint = {1009.1617},
 primaryClass = {astro-ph.GA},
       adsurl = {https://ui.adsabs.harvard.edu/abs/2010ApJ...723L...7K}
}

@ARTICLE{2013A&A...552A..40C,
       author = {{Chira}, R.-A. and {Beuther}, H. and {Linz}, H. and {Schuller}, F. and {Walmsley}, C.~M. and {Menten}, K.~M. and {Bronfman}, L.},
        title = "{Characterization of infrared dark clouds. NH$_{3}$ observations of an absorption-contrast selected IRDC sample}",
      journal = {\aap},
         year = 2013,
        month = apr,
       volume = {552},
          eid = {A40},
        pages = {A40},
          doi = {10.1051/0004-6361/201219567},
archivePrefix = {arXiv},
       eprint = {1302.6774},
 primaryClass = {astro-ph.SR},
       adsurl = {https://ui.adsabs.harvard.edu/abs/2013A&A...552A..40C}
}

@ARTICLE{2014MNRAS.439.3275W,
       author = {{Wang}, Ke and {Zhang}, Qizhou and {Testi}, Leonardo and {van der Tak}, Floris and {Wu}, Yuefang and {Zhang}, Huawei and {Pillai}, Thushara and {Wyrowski}, Friedrich and {Carey}, Sean and {Ragan}, Sarah E. and et al.},
        title = "{Hierarchical fragmentation and differential star formation in the Galactic `Snake': infrared dark cloud G11.11-0.12}",
      journal = {\mnras},
         year = 2014,
        month = apr,
       volume = {439},
       number = {4},
        pages = {3275-3293},
          doi = {10.1093/mnras/stu127},
archivePrefix = {arXiv},
       eprint = {1401.4157},
 primaryClass = {astro-ph.GA},
       adsurl = {https://ui.adsabs.harvard.edu/abs/2014MNRAS.439.3275W}
}

@ARTICLE{2006A&A...447..929P,
       author = {{Pillai}, T. and {Wyrowski}, F. and {Menten}, K.~M. and {Kr{\"u}gel}, E.},
        title = "{High mass star formation in the infrared dark cloud G11.11-0.12}",
      journal = {\aap},
         year = 2006,
        month = mar,
       volume = {447},
       number = {3},
        pages = {929-936},
          doi = {10.1051/0004-6361:20042145},
archivePrefix = {arXiv},
       eprint = {astro-ph/0510622},
 primaryClass = {astro-ph},
       adsurl = {https://ui.adsabs.harvard.edu/abs/2006A&A...447..929P}
}

@ARTICLE{2002ApJ...566..945B,
       author = {{Beuther}, H. and {Schilke}, P. and {Menten}, K.~M. and {Motte}, F. and {Sridharan}, T.~K. and {Wyrowski}, F.},
        title = "{High-Mass Protostellar Candidates. II. Density Structure from Dust Continuum and CS Emission}",
      journal = {\apj},
         year = 2002,
        month = feb,
       volume = {566},
       number = {2},
        pages = {945-965},
          doi = {10.1086/338334},
archivePrefix = {arXiv},
       eprint = {astro-ph/0110370},
 primaryClass = {astro-ph},
       adsurl = {https://ui.adsabs.harvard.edu/abs/2002ApJ...566..945B}
}

@ARTICLE{2006ApJ...641..389R,
       author = {{Rathborne}, J.~M. and {Jackson}, J.~M. and {Simon}, R.},
        title = "{Infrared Dark Clouds: Precursors to Star Clusters}",
      journal = {\apj},
         year = 2006,
        month = apr,
       volume = {641},
       number = {1},
        pages = {389-405},
          doi = {10.1086/500423},
archivePrefix = {arXiv},
       eprint = {astro-ph/0602246},
 primaryClass = {astro-ph},
       adsurl = {https://ui.adsabs.harvard.edu/abs/2006ApJ...641..389R}
}

@ARTICLE{2009A&A...505..405P,
       author = {{Peretto}, N. and {Fuller}, G.~A.},
        title = "{The initial conditions of stellar protocluster formation. I. A catalogue of Spitzer dark clouds}",
      journal = {\aap},
         year = 2009,
        month = oct,
       volume = {505},
       number = {1},
        pages = {405-415},
          doi = {10.1051/0004-6361/200912127},
archivePrefix = {arXiv},
       eprint = {0906.3493},
 primaryClass = {astro-ph.GA},
       adsurl = {https://ui.adsabs.harvard.edu/abs/2009A&A...505..405P}
}

@ARTICLE{2017MNRAS.469.2163E,
       author = {{Eden}, D.~J. and {Moore}, T.~J.~T. and {Plume}, R. and {Urquhart}, J.~S. and {Thompson}, M.~A. and {Parsons}, H. and {Dempsey}, J.~T. and {Rigby}, A.~J. and {Morgan}, L.~K. and {Thomas}, H.~S. and et al.},
        title = "{The JCMT Plane Survey: first complete data release - emission maps and compact source catalogue}",
      journal = {\mnras},
         year = 2017,
        month = aug,
       volume = {469},
       number = {2},
        pages = {2163-2183},
          doi = {10.1093/mnras/stx874},
archivePrefix = {arXiv},
       eprint = {1704.02982},
 primaryClass = {astro-ph.GA},
       adsurl = {https://ui.adsabs.harvard.edu/abs/2017MNRAS.469.2163E}
}

@ARTICLE{2019MNRAS.485.2895E,
       author = {{Eden}, D.~J. and {Liu}, Tie and {Kim}, Kee-Tae and {Juvela}, M. and {Liu}, S.-Y. and {Tatematsu}, K. and {Francesco}, J. Di and {Wang}, K. and {Wu}, Y. and {Thompson}, M.~A. and et al.},
        title = "{SCOPE: SCUBA-2 Continuum Observations of Pre-protostellar Evolution - survey description and compact source catalogue}",
      journal = {\mnras},
         year = 2019,
        month = may,
       volume = {485},
       number = {2},
        pages = {2895-2908},
          doi = {10.1093/mnras/stz574},
archivePrefix = {arXiv},
       eprint = {1902.10180},
 primaryClass = {astro-ph.GA},
       adsurl = {https://ui.adsabs.harvard.edu/abs/2019MNRAS.485.2895E}
}

@ARTICLE{2016MNRAS.457.2675H,
       author = {{Henshaw}, J.~D. and {Longmore}, S.~N. and {Kruijssen}, J.~M.~D. and {Davies}, B. and {Bally}, J. and {Barnes}, A. and {Battersby}, C. and {Burton}, M. and {Cunningham}, M.~R. and {Dale}, J.~E. and {Ginsburg}, A. and {Immer}, K. and {Jones}, P.~A. and {Kendrew}, S. and {Mills}, E.~A.~C. and {Molinari}, S. and {Moore}, T.~J.~T. and {Ott}, J. and {Pillai}, T. and {Rathborne}, J. and {Schilke}, P. and {Schmiedeke}, A. and {Testi}, L. and {Walker}, D. and {Walsh}, A. and {Zhang}, Q.},
        title = "{Molecular gas kinematics within the central 250 pc of the Milky Way}",
      journal = {\mnras},
         year = 2016,
        month = apr,
       volume = {457},
       number = {3},
        pages = {2675-2702},
          doi = {10.1093/mnras/stw121},
archivePrefix = {arXiv},
       eprint = {1601.03732},
 primaryClass = {astro-ph.GA},
       adsurl = {https://ui.adsabs.harvard.edu/abs/2016MNRAS.457.2675H}
}

@ARTICLE{2019MNRAS.485.2457H,
       author = {{Henshaw}, J.~D. and {Ginsburg}, A. and {Haworth}, T.~J. and {Longmore}, S.~N. and {Kruijssen}, J.~M.~D. and {Mills}, E.~A.~C. and {Sokolov}, V. and {Walker}, D.~L. and {Barnes}, A.~T. and {Contreras}, Y. and et al.},
        title = "{`The Brick' is not a brick: a comprehensive study of the structure and dynamics of the central molecular zone cloud G0.253+0.016}",
      journal = {\mnras},
         year = 2019,
        month = may,
       volume = {485},
       number = {2},
        pages = {2457-2485},
          doi = {10.1093/mnras/stz471},
archivePrefix = {arXiv},
       eprint = {1902.02793},
 primaryClass = {astro-ph.GA},
       adsurl = {https://ui.adsabs.harvard.edu/abs/2019MNRAS.485.2457H}
}

@ARTICLE{2013A&A...554A..55H,
       author = {{Hacar}, A. and {Tafalla}, M. and {Kauffmann}, J. and {Kov{\'a}cs}, A.},
        title = "{Cores, filaments, and bundles: hierarchical core formation in the L1495/B213 Taurus region}",
      journal = {\aap},
         year = 2013,
        month = jun,
       volume = {554},
          eid = {A55},
        pages = {A55},
          doi = {10.1051/0004-6361/201220090},
archivePrefix = {arXiv},
       eprint = {1303.2118},
 primaryClass = {astro-ph.GA},
       adsurl = {https://ui.adsabs.harvard.edu/abs/2013A&A...554A..55H}
}

@ARTICLE{2023ApJ...944L..15S,
       author = {{Schinnerer}, Eva and {Emsellem}, Eric and {Henshaw}, Jonathan D. and {Liu}, Daizhong and {Meidt}, Sharon E. and {Querejeta}, Miguel and {Renaud}, Florent and {Sormani}, Mattia C. and {Sun}, Jiayi and {Egorov}, Oleg V. and et al.},
        title = "{PHANGS-JWST First Results: Rapid Evolution of Star Formation in the Central Molecular Gas Ring of NGC 1365}",
      journal = {\apjl},
         year = 2023,
        month = feb,
       volume = {944},
       number = {2},
          eid = {L15},
        pages = {L15},
          doi = {10.3847/2041-8213/acac9e},
archivePrefix = {arXiv},
       eprint = {2212.09168},
 primaryClass = {astro-ph.GA},
       adsurl = {https://ui.adsabs.harvard.edu/abs/2023ApJ...944L..15S}
}

@ARTICLE{2008ApJ...679.1338R,
       author = {{Rosolowsky}, E.~W. and {Pineda}, J.~E. and {Kauffmann}, J. and {Goodman}, A.~A.},
        title = "{Structural Analysis of Molecular Clouds: Dendrograms}",
      journal = {\apj},
         year = 2008,
        month = jun,
       volume = {679},
       number = {2},
        pages = {1338-1351},
          doi = {10.1086/587685},
archivePrefix = {arXiv},
       eprint = {0802.2944},
 primaryClass = {astro-ph},
       adsurl = {https://ui.adsabs.harvard.edu/abs/2008ApJ...679.1338R}
}

@ARTICLE{2019ApJS..240....9S,
       author = {{Su}, Yang and {Yang}, Ji and {Zhang}, Shaobo and {Gong}, Yan and {Wang}, Hongchi and {Zhou}, Xin and {Wang}, Min and {Chen}, Zhiwei and {Sun}, Yan and {Chen}, Xuepeng and et al.},
        title = "{The Milky Way Imaging Scroll Painting (MWISP): Project Details and Initial Results from the Galactic Longitudes of 25.{\textdegree}8-49.{\textdegree}7}",
      journal = {\apjs},
         year = 2019,
        month = jan,
       volume = {240},
       number = {1},
          eid = {9},
        pages = {9},
          doi = {10.3847/1538-4365/aaf1c8},
archivePrefix = {arXiv},
       eprint = {1901.00285},
 primaryClass = {astro-ph.GA},
       adsurl = {https://ui.adsabs.harvard.edu/abs/2019ApJS..240....9S}
}

@ARTICLE{2019ApJ...885..131R,
       author = {{Reid}, M.~J. and {Menten}, K.~M. and {Brunthaler}, A. and {Zheng}, X.~W. and {Dame}, T.~M. and {Xu}, Y. and {Li}, J. and {Sakai}, N. and {Wu}, Y. and {Immer}, K. and {Zhang}, B. and {Sanna}, A. and {Moscadelli}, L. and {Rygl}, K.~L.~J. and {Bartkiewicz}, A. and {Hu}, B. and {Quiroga-Nu{\~n}ez}, L.~H. and {van Langevelde}, H.~J.},
        title = "{Trigonometric Parallaxes of High-mass Star-forming Regions: Our View of the Milky Way}",
      journal = {\apj},
         year = 2019,
        month = nov,
       volume = {885},
       number = {2},
          eid = {131},
        pages = {131},
          doi = {10.3847/1538-4357/ab4a11},
archivePrefix = {arXiv},
       eprint = {1910.03357},
 primaryClass = {astro-ph.GA},
       adsurl = {https://ui.adsabs.harvard.edu/abs/2019ApJ...885..131R}
}

@BOOK{1990sse..book.....K,
       author = {{Kippenhahn}, Rudolf and {Weigert}, Alfred},
        title = "{Stellar Structure and Evolution}",
         year = 1990,
       adsurl = {https://ui.adsabs.harvard.edu/abs/1990sse..book.....K}
}

@ARTICLE{1956MNRAS.116..351B,
       author = {{Bonnor}, W.~B.},
        title = "{Boyle's Law and gravitational instability}",
      journal = {\mnras},
         year = 1956,
        month = jan,
       volume = {116},
        pages = {351},
          doi = {10.1093/mnras/116.3.351},
       adsurl = {https://ui.adsabs.harvard.edu/abs/1956MNRAS.116..351B}
}

@ARTICLE{1983ApJ...271..417F,
       author = {{Frenk}, C.~S. and {White}, S.~D.~M. and {Davis}, M.},
        title = "{Nonlinear evolution of large-scale structure in the universe}",
      journal = {\apj},
         year = 1983,
        month = aug,
       volume = {271},
        pages = {417-430},
          doi = {10.1086/161209},
       adsurl = {https://ui.adsabs.harvard.edu/abs/1983ApJ...271..417F}
}

@ARTICLE{2005ApJ...625..891R,
       author = {{Reid}, Michael A. and {Wilson}, Christine D.},
        title = "{High-Mass Star Formation. I. The Mass Distribution of Submillimeter Clumps in NGC 7538}",
      journal = {\apj},
         year = 2005,
        month = jun,
       volume = {625},
       number = {2},
        pages = {891-905},
          doi = {10.1086/429790},
archivePrefix = {arXiv},
       eprint = {astro-ph/0503190},
 primaryClass = {astro-ph},
       adsurl = {https://ui.adsabs.harvard.edu/abs/2005ApJ...625..891R}
}

@ARTICLE{2007ApJ...655..351L,
       author = {{Li}, D. and {Velusamy}, T. and {Goldsmith}, P.~F. and {Langer}, William D.},
        title = "{Massive Quiescent Cores in Orion. II. Core Mass Function}",
      journal = {\apj},
         year = 2007,
        month = jan,
       volume = {655},
       number = {1},
        pages = {351-363},
          doi = {10.1086/509736},
archivePrefix = {arXiv},
       eprint = {astro-ph/0610634},
 primaryClass = {astro-ph},
       adsurl = {https://ui.adsabs.harvard.edu/abs/2007ApJ...655..351L}
}

@ARTICLE{2019ApJ...886..102S,
       author = {{Sanhueza}, Patricio and {Contreras}, Yanett and {Wu}, Benjamin and {Jackson}, James M. and {Guzm{\'a}n}, Andr{\'e}s E. and {Zhang}, Qizhou and {Li}, Shanghuo and {Lu}, Xing and {Silva}, Andrea and {Izumi}, Natsuko and et al.},
        title = "{The ALMA Survey of 70 {\ensuremath{\mu}}m Dark High-mass Clumps in Early Stages (ASHES). I. Pilot Survey: Clump Fragmentation}",
      journal = {\apj},
         year = 2019,
        month = dec,
       volume = {886},
       number = {2},
          eid = {102},
        pages = {102},
          doi = {10.3847/1538-4357/ab45e9},
archivePrefix = {arXiv},
       eprint = {1909.07985},
 primaryClass = {astro-ph.GA},
       adsurl = {https://ui.adsabs.harvard.edu/abs/2019ApJ...886..102S}
}

@ARTICLE{2013A&A...558A..33A,
       author = {{Astropy Collaboration} and {Robitaille}, Thomas P. and {Tollerud}, Erik J. and {Greenfield}, Perry and {Droettboom}, Michael and {Bray}, Erik and {Aldcroft}, Tom and {Davis}, Matt and {Ginsburg}, Adam and {Price-Whelan}, Adrian M. and et al.},
        title = "{Astropy: A community Python package for astronomy}",
      journal = {\aap},
         year = 2013,
        month = oct,
       volume = {558},
          eid = {A33},
        pages = {A33},
          doi = {10.1051/0004-6361/201322068},
archivePrefix = {arXiv},
       eprint = {1307.6212},
 primaryClass = {astro-ph.IM},
       adsurl = {https://ui.adsabs.harvard.edu/abs/2013A&A...558A..33A}
}

@ARTICLE{2018AJ....156..123A,
       author = {{Astropy Collaboration} and {Price-Whelan}, A.~M. and {Sip{\H{o}}cz}, B.~M. and {G{\"u}nther}, H.~M. and {Lim}, P.~L. and {Crawford}, S.~M. and {Conseil}, S. and {Shupe}, D.~L. and {Craig}, M.~W. and {Dencheva}, N. and et al.},
        title = "{The Astropy Project: Building an Open-science Project and Status of the v2.0 Core Package}",
      journal = {\aj},
         year = 2018,
        month = sep,
       volume = {156},
       number = {3},
          eid = {123},
        pages = {123},
          doi = {10.3847/1538-3881/aabc4f},
archivePrefix = {arXiv},
       eprint = {1801.02634},
 primaryClass = {astro-ph.IM},
       adsurl = {https://ui.adsabs.harvard.edu/abs/2018AJ....156..123A}
}

@ARTICLE{2022ApJ...935..167A,
       author = {{Astropy Collaboration} and {Price-Whelan}, Adrian M. and {Lim}, Pey Lian and {Earl}, Nicholas and {Starkman}, Nathaniel and {Bradley}, Larry and {Shupe}, David L. and {Patil}, Aarya A. and {Corrales}, Lia and {Brasseur}, C.~E. and et al.},
        title = "{The Astropy Project: Sustaining and Growing a Community-oriented Open-source Project and the Latest Major Release (v5.0) of the Core Package}",
      journal = {\apj},
         year = 2022,
        month = aug,
       volume = {935},
       number = {2},
          eid = {167},
        pages = {167},
          doi = {10.3847/1538-4357/ac7c74},
archivePrefix = {arXiv},
       eprint = {2206.14220},
 primaryClass = {astro-ph.IM},
       adsurl = {https://ui.adsabs.harvard.edu/abs/2022ApJ...935..167A}
}

@ARTICLE{1974ITAC...19..716A,
       author = {{Akaike}, H.},
        title = "{A New Look at the Statistical Model Identification}",
      journal = {IEEE Transactions on Automatic Control},
         year = 1974,
        month = jan,
       volume = {19},
        pages = {716-723},
          doi = {10.1109/TAC.1974.1100705},
       adsurl = {https://ui.adsabs.harvard.edu/abs/1974ITAC...19..716A}
}
\bibliographystyle{aasjournalv7}



\end{document}